# Distortion-Driven Carrier Decoupling in Doped LiMgPO$_4$

Zhihua Zheng[1], Xiaolong Yao[1,4,*], Cailian Yu[1], Menghao Gao[1], Fangping Ouyang[1,2,3,*], Shiwu Gao[4,*]

[1]*School of Physical Science and Technology, Xinjiang Key Laboratory of Solid-State Physics and Devices, Xinjiang University, Urumqi 830017, China*

[2]*School of Physics, Institute of Quantum Physics, Hunan Key Laboratory for Super-Microstructure and Ultrafast Process, and Hunan Key Laboratory of Nanophotonics and Devices, Central South University, Changsha 410083, China*

[3]*State Key Laboratory of Powder Metallurgy, and Powder Metallurgy Research Institute, Central South University, Changsha 410083, People's Republic of China*

[4]*Beijing Computational Science Research Center, Beijing 100193, China*

[*]*Email: xlyao@xju.edu.cn, ouyangfp06@tsinghua.org.cn, swgao@csrc.ac.cn*

**Abstract**

The interplay between lattice distortions and charge carriers governs the properties of many functional oxides. In alkali-doped LiMgPO$_4$, a significant enhancement in dosimetric response is observed, but its microscopic origin is not understood. Using non-adiabatic molecular dynamics, we reveal a fundamental mechanism of carrier decoupling driven by a hierarchy of lattice distortions. We show that electrons localize into stable small polarons on an ultrafast timescale, trapped by the strong local potential induced by the dopant, while holes form more delocalized polarons that migrate efficiently through a lattice smoothed by global strain. The stark contrast between the dynamics of trapped electrons and mobile holes explains the suppressed recombination and enhanced energy storage. These results present a clear physical picture of how multiscale lattice distortions can independently control electron and hole transport, offering new insights into the physics of polarons in complex materials.

**TOC Graphic**

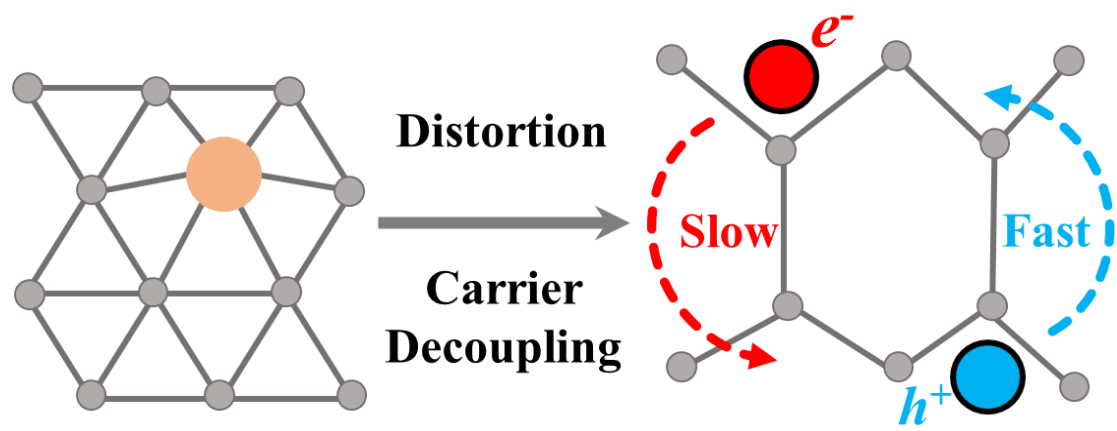

Stimulated luminescence - the emission of light from an irradiated crystalline solid upon thermal or optical excitation - forms the basis of modern radiation dosimetry.[1-3] While established detector materials such as $Al_2O_3$ and BeO are commercially available, they often suffer from high manufacturing costs or suboptimal trade-offs in performance metrics,[4-7] driving the search for superior alternatives. Among them, metal phosphates have garnered significant attention for their exceptional microwave dielectric properties, thermal stability, and pronounced radiation sensitivity, making them highly valuable for dosimetry applications.[8-10] This has spurred progressively deeper investigations into phosphate compounds in recent years.[11,12] In particular, $LiMgPO_4$ has emerged as a promising candidate, offering a balanced performance profile.[13-15] However, for it to surpass commercial standards, two critical bottlenecks must be addressed: its relatively low luminescence efficiency and rapid signal fading.

Experimental strategies to overcome these limitations have historically focused on rare-earth doping, [16-19] but a recent breakthrough by Kalinkin *et al.* demonstrated that isovalent Na doping can also markedly enhance thermoluminescence output and modify absorption,[20] merely 6 % Na substitution boosts the thermoluminescence output markedly, red-shifts the UV–Vis absorption edge and monotonically increases the ESR intensity. Despite this rapid experimental progress, a fundamental theoretical understanding of how alkali metal doping simultaneously amplifies emission and suppresses fading remains nascent. Current theoretical investigations of $LiMgPO_4$ are not only scarce but also suffer from fundamental limitations. The few existing studies, which attempt to explain performance variations through oxygen vacancies or aliovalent defects, rely heavily on static density functional theory (DFT) calculations of the electronic band structure.[21-23] This approach fails to capture crucial features such as resonant defect states within the conduction and valence bands, which govern charge carrier trapping and recombination pathways. More significantly, these models largely neglect the strong electron-phonon coupling inherent in such ionic crystals. Consequently, they cannot adequately describe the dynamic local lattice distortions that trap charge carriers to form polarons - a process fundamental to carrier mobility and, ultimately, the dosimetric properties of the material. This disconnect between experimental discovery and fundamental theory now constitutes a key barrier to transformative progress.

To bridge this gap, this work employs a multiscale theoretical framework combining static and dynamic simulations. We first use first-principles calculations to systematically analyze the geometric structures, electronic properties, and optical responses of $LiMgPO_4$ doped with Na, K, and Rb. This static analysis identifies critical defect-induced states near the band edges. Crucially, we then employ *ab initio* nonadiabatic molecular dynamics (NAMD) to quantitatively characterize the real-time, non-equilibrium dynamics of charge carriers.[24] By dissecting the interplay between dopant-induced lattice distortions and carrier behavior, we reveal the microscopic mechanisms governing performance. Our findings demonstrate that alkali metal doping synergistically modulates carrier dynamics: localized lattice distortions create self-trapping potentials that slow electron relaxation, thereby enhancing energy storage, while long-range strain facilitates hole migration, suppressing detrimental carrier recombination. This work resolves two critical omissions in prior theoretical models - the role of resonant defect states and the impact of polaron dynamics - and provides a comprehensive, mechanism-based explanation for the observed performance enhancements. The integrated static-dynamic approach presented here establishes a robust new paradigm for the theory-guided design of advanced radiation dosimetry materials.

Our investigation begins with the atomic and electronic structure of LiMgPO$_4$, modeled using a 1×2×1 supercell (56 atoms, space group Pnma) as depicted in Figure 1a. The relaxed lattice parameters of the pristine structure ($a$ = 4.73 Å, $b$ = 5.96 Å, $c$ = 10.24 Å) are in excellent agreement with experimental values (deviation <1%).[21,22] To quantify the doping-induced lattice distortion, we select the bond lengths ($d_1$ and $d_2$) and the bond angle ($\theta$) between the substituent site and neighboring oxygen atoms as primary structural descriptors (Figure 1b).

Motivated by experimental reports of enhanced thermoluminescence in Na-doped LiMgPO$_4$ by Kalinkin et al.,[20] our investigation extends this doping strategy to homologous alkali metals (K, Rb). The substitution of Li$^+$ with larger alkali metal cations (Na$^+$, K$^+$, Rb$^+$) induces a systematic lattice distortion, the magnitude of which correlates directly with the dopant's ionic radius (Figure 1c). For instance, in the Rb-doped system, the $d_1$ and $d_2$ bonds expand from 2.11 Å and 2.19 Å to 2.60 Å and 2.60 Å, respectively, representing increases of 23.2% and 18.7%. Concurrently, the bond angle $\theta$ contracts from 90.12° to 82.11°, an 8.9% decrease. This pronounced local deformation around the dopant site generates a long-range strain field, resulting in an anisotropic global expansion of the supercell along all three crystallographic axes (Figure 1d, Tables S1 and S3). The monotonic increase of lattice parameters $a$, $b$, and $c$ with dopant size observed in our simulations (Table S2) mirrors exactly the trend reported in the experimental XRD data of Kalinkin *et al.* (their Table 1), validating that the dopant-induced distortion is faithfully captured by our structural models. The most significant effect is observed with Rb doping, which increases the supercell volume by 4.8% (from 576.82 Å$^3$ to 604.55 Å$^3$).

To assess the thermodynamic stability of these heavily distorted structures, particularly for multi-dopant configurations (Figure S1a), we perform *ab initio* molecular dynamics (AIMD) simulations within the NVT ensemble. Even the triple-doped system remains structurally stable throughout a 15-ps simulation at 300 K, exhibiting minimal energy fluctuations (<0.03 eV/atom) and no bond dissociation (Figure S1b). The integrity of the PO$_4$ tetrahedra is preserved, confirming that the fundamental crystal framework withstands the induced strain.

Defect formation energy ($E_f$) calculations further corroborate the thermodynamic feasibility of this doping strategy. Under Li-poor conditions, all single-dopant configurations exhibit negative formation energies, indicating they are thermodynamically more favorable than intrinsic Li vacancies ($V_{Li}$). While $E_f$ increases with dopant size [$E_f$(Rb) > $E_f$(K) > $E_f$(Na)], the formation energies for double- and triple-doped systems are even more favorable, suggesting a thermodynamic driving force for achieving high dopant concentrations (Figure S2).

Having established the stability of these dopant-induced lattice distortions, we proceeded to examine their profound impact on the electronic structure. We hypothesized that these atomic-scale structural modifications are the key to modulating the electronic properties and subsequent carrier dynamics. This is confirmed by the systematic reduction of the band gap ($E_g$) with increasing dopant ionic radius and volumetric expansion ($\Delta V/V_0$), as shown in Figure 2b-d and Table 1. While the PBE functional underestimates the absolute band gap value (5.47 eV[22] for the pristine system compared to the experimental optical gap of ~7.5 eV[25]), this well-known limitation does not affect the qualitative trends induced by doping. Our calculations reveal a clear progression: the gap narrows to 5.34 eV (Na, +1.39% $\Delta V/V_0$), 5.10 eV (K, +3.62% $\Delta V/V_0$), and ultimately to 4.92 eV (Rb, +4.80% $\Delta V/V_0$). Critically, this band gap reduction is accompanied by the formation of new, shallow donor-like states near the conduction band minimum (CBM) at the Γ-point.

In stark contrast, the intrinsic lithium vacancy ($V_{Li}$), despite causing a modest lattice expansion (+0.45% $\Delta V/V_0$), widens the band gap to 5.52 eV. This finding demonstrates that volumetric strain is not the sole determinant of the band gap. Furthermore, $V_{Li}$ introduces a distinct, localized acceptor-like state just above the valence band maximum (VBM) (Figure S3d). Analysis of the partial density of states (PDOS) in Figure S4b confirms that this state originates from the unsaturated $2p$ orbitals of the undercoordinated oxygen atoms neighboring the vacancy, creating a potential hole-trapping center.

Multi-dopant configurations further underscore the nuanced interplay between strain and chemistry. For instance, Na-K co-doping induces a substantial 4.90% volume expansion, comparable to that of single Rb-doping, yet yields only a modest bandgap of 5.00 eV (Table S2 and Figure S3). This decoupling of strain and $E_g$ highlights that the electronic structure is sensitively modulated by the specific chemical identity and concentration of the dopants, not just the resulting lattice deformation. The triple-doped (Na, K, Rb) system exhibits the strongest synergistic effect, achieving both the largest lattice expansion (+14.91%) and the narrowest band gap (4.78 eV), along with the most prominent emergence of states near the CBM.

The orbital origins of these electronic modifications are revealed by the PDOS (Figure S4). While the VBM remains dominated by O 2$p$ orbitals in all systems, the nature of the CBM is substantially altered. In doped systems, the CBM is reconstructed through a hybridization of the host Li and Mg $s$-orbitals with the vacant $s$-orbitals of the alkali dopant atoms. This reconstruction, driven by both the local chemical environment and the long-range strain, creates the aforementioned shallow electronic states. The fundamentally ionic character of the bonding, confirmed by the electron localization function (ELF) analysis (Figure S5), remains intact across all doped systems.

Our electronic structure analysis reveals that alkali-metal doping engineers a tailored landscape of localized defect states near the band edges. The dopants create shallow electron-trapping states at the CBM, while intrinsic defects like $V_{Li}$ create deep hole-trapping states at the VBM. To directly probe how this engineered electronic terrain dictates the charge carrier dynamics, specifically the formation, localization, and transport of electron and hole polarons, we turned to nonadiabatic calculations. These simulations were performed on structural configurations sampled from ab initio molecular dynamics (AIMD) simulations at 300 K, which inherently embed the thermal fluctuations crucial for activating phonon-assisted quantum transitions.[26]

The temporal evolution of electron energy levels within the five lowest conduction bands (CBM to CBM+4) at the Γ-point reveals the influence of doping on relaxation dynamics (Figure 3a-d). In pristine LiMgPO$_4$, the electron population excited to higher conduction bands relaxes, with 50% of the energy transferred back towards the CBM within 1000 fs and full relaxation occurring by approximately 2500 fs. Na-doping accelerates this process, reducing the 50% relaxation time to 800 fs - a 20% decrease compared to the pristine system. In stark contrast, K- and Rb-doping significantly hinder electron relaxation. K-doping extends the 50% relaxation time to 1800 fs (an 80% increase), while Rb-doping extends it to 1200 fs (a 20% increase). For both K- and Rb-doped systems, complete relaxation is delayed to beyond 3000 fs, indicating the formation of more stable, localized electron states.

Hole dynamics within the five highest valence bands (VBM to VBM-4) exhibit a contrasting trend (Figure 3e-h). In pristine LiMgPO$_4$, holes rapidly self-trap at the VBM within 500 fs, forming localized hole polarons that remain largely immobile. Upon doping, this initial localization accelerates to under 300 fs, but is immediately followed by enhanced quantum hopping. Specifically, the Na-doped system facilitates two distinct VBM↔VBM-1 transitions. This effect is even more pronounced in the K-doped system, which displays three frequent hopping events. The Rb-doped system not only shows two VBM ↔ VBM-1 transitions but also transiently populates the deeper

VBM-2 level, indicating access to additional transport pathways.

This synergistic interplay between electron and hole dynamics explains the enhanced radiation detection performance. Efficient hole polaron migration, promoted by doping, effectively separates the charge carriers and suppresses their recombination at the initial excitation sites. Concurrently, the prolonged lifetime of the localized, self-trapped electrons (STEs) in the K- and Rb-doped systems facilitates long-term energy storage and signal stability. This framework clarifies why Na-doping, despite its accelerated electron relaxation, yields superior performance: its substantially enhanced hole mobility provides a critical compensatory advantage. Meanwhile, K-doped $LiMgPO_4$ achieves an optimal balance: its significantly retarded electron relaxation (50% decay time of 1800 fs) is coupled with highly efficient hole transport (three hopping events), leading to a combined enhancement of its detection capabilities.

Analysis of time-dependent carrier localization provides crucial insight into the nature of the charge carriers. We first examine the electron dynamics, which are dominated by the local trapping environment. Electron localization within the dopant-centered defect fragment-the substitutional dopant and its six nearest-neighbor oxygens (highlighted in Figure 1b)-mirrors the evolution trend shown in Figure 3a-d. In pristine $LiMgPO_4$ the fraction of electron density confined to this fragment decays uniformly from 45 % to 40 % over 3 ps (Figure 4a), consistent with the rapid 1000 fs energy relaxation in Fig. 3. Na-doping accelerates both processes: localization drops to 40 % within ~1500 fs (Figure 4b) and the 50 % energy relaxation shortens to 1000 fs (Figure 3a-d). In contrast, K- and Rb-doped systems slow the descent: localization barely decreases and remains locked at ~40 % after 1 ps (Figure 4c-d), paralleling the extended 50 % relaxation times of 1800 fs (K) and 1200 fs (Rb) seen in Figure 3a-d. This synchronized slowing signals the formation of stable small electron polarons whose lifetime is directly controlled by the local distortion introduced by the larger alkali cation.

Having established how local environment governs electron behavior, we now clarify the hole dynamics by distinguishing the local character of the polaron's wavefunction from the global mobility of the polaron as a quasi-particle. Figure 4 reveals the hole's wavefunction core remains strongly anchored, with approximately 75% of its charge density consistently localized on the same dopant-centered defect fragment. This describes the internal structure of a stable hole polaron at a given site. However, this entire quasi-particle - the hole dressed by its local lattice distortion - is mobile. It can tunnel between crystallographically equivalent sites, a process seen as the enhanced quantum hopping in the energy dynamics of Figure 3e-h. This resolves the apparent contradiction: the polaron is locally well-defined and stable (anchored) but globally mobile as it hops through the crystal lattice. This is in stark contrast to the electron dynamics, which are dominated by the local trapping environment. This mechanism-based decoupling provides a powerful strategy for materials design: hole polaron mobility can be tuned via global lattice engineering, while electron polaron lifetimes can be independently controlled by targeted doping at the defect site.

*Ab initio* NAMD simulations provide a quantum-mechanical rationale for these decoupled dynamics (Figure 5). In the valence band region, doping with larger alkali metals progressively intensifies energy-level crossings and fluctuations. This creates an increasingly dense network of transient states (Figure 5b-d, blue regions), which strengthens the electron-phonon coupling and establishes a "quantum highway" for efficient hole polaron tunneling. In the conduction band, the trend is reversed. Larger dopants (K, Rb) suppress level crossings near the CBM, creating sparser state distributions (Figure 5c-d, red regions). This weakened nonadiabatic coupling hinders thermal relaxation pathways and effectively creates a localization trap for electrons. This suppression is further amplified by an increased energy gap between the CBM and CBM+1 levels, which reduces the probability of nonadiabatic transitions. Notably, this targeted modification of the conduction band edge structure occurs while the fundamental band gap of the material remains largely unchanged, preserving its primary optical properties.[27] The magnitude of this suppression (K > Rb > Na) correlates directly with the observed increase in electron relaxation times. This synergy between static distortion and dynamic fluctuations thus gives rise to two distinct quantum transport regimes: a delocalized hole polaron transport channel and a localized electron polaron trap, which can be orthogonally engineered to optimize material performance.

To quantitatively validate this mechanism, we calculated the time-averaged root-mean-square non-adiabatic coupling (NAC) values, which govern the transition probabilities between electronic states (Figure 5e-f). The results provide direct evidence for the decoupled dynamics. For electrons, as shown in Figure 5e, the NAC between the CBM and CBM+4 shows an overall decreasing trend with the progression of the doping element. This suppression of the nonadiabatic coupling pathway hinders electron relaxation and explains the formation of the localization trap observed in the dynamics. Conversely, for holes, as shown in Figure 5f, the NAC between the VBM and VBM-4 shows an overall increasing trend with the progression of the doping element. This significant enhancement of the coupling strength confirms the formation of a "quantum highway" that facilitates efficient hopping and transport. More detailed data are shown in Figures S6 and S7. This quantitative analysis unequivocally demonstrates that alkali-metal doping differentially modulates the electron-phonon coupling for electrons and holes, providing a robust physical basis for the observed synergistic carrier dynamics. We emphasize that this enhanced hole mobility is an intrinsic property driven by the dopant-induced modifications to the valence band's non-adiabatic coupling (Figure 5f). It is not an apparent delocalization caused merely by the electron's separate localization. Rather, these two effects: (1) electrons being trapped by local distortions and (2) holes gaining mobility from global strain - are parallel, synergistic mechanisms that both contribute to suppressing carrier recombination and enhancing the material's performance, which ultimately manifests in the material's superior macroscopic optical properties.

Alkali metal doping enhances the optical response of $LiMgPO_4$ through two distinct mechanisms rooted in lattice distortion. First, doping modifies the electronic band structure, reducing the band gap from 5.47 eV in the pristine host to facilitate interband transitions. Second, the dopants and the associated local lattice distortions introduce defect states near the CBM. These states can act as shallow traps, effectively localizing photoexcited electrons and prolonging their relaxation lifetime, a mechanism crucial for signal stability. These effects are directly manifested in the optical absorption spectra (Figure S8), which show a systematic redshift and an overall increase in absorption intensity upon doping. Furthermore, the absorption enhancement for various systems, including co-doped and vacancy-containing configurations, is broadly correlated with the presence of dopants and lattice distortion.

A quantitative analysis of the single-dopant systems reveals a more complex interplay between these mechanisms. At a wavelength of 180 nm, Rb-doping nearly doubles the light absorption intensity compared to the undoped host, with the overall enhancement following the order Rb > K > Na. This calculated systematic redshift of the absorption edge (shown in Figure S8a) is in excellent qualitative agreement with experimental measurements. Specifically, Kalinkin *et al.*[20] (see Figure 8 in their work) reported a clear, gradual red-shift of the UV-Vis absorption edge with increasing Na content in $Li_{1-x}Na_xMgPO_4$. Our calculations, which focus on the effect of the substitutional dopant and its associated lattice strain, capture this fundamental trend. This strongly indicates that the dopant-induced lattice distortion is a non-negligible physical mechanism contributing to the observed red-shift, even though the full magnitude of the experimental shift is likely also influenced by other factors, such as the defect states (e.g., oxygen vacancies) that accompany the doping process. This trend correlates directly with the calculated lattice distortion amplitude ($\Delta V/V$: 4.8% for Rb, 3.6% for K, 1.4% for Na). Interestingly, the absorption enhancement is inversely correlated with the magnitude of the band gap reduction ($\Delta E_g/E_g$: -10.1% for Na, -6.8% for K, -2.4% for Rb). This key finding suggests that the local lattice distortion and the consequent introduction of efficient carrier-localizing states are the dominant factor driving the optical absorption enhancement, outweighing the influence of the global band gap reduction.

Analysis of the PDOS further illuminates these electronic structure modifications. The primary absorption edge in pristine $LiMgPO_4$ is attributed to O $2p$ → Mg $3p$ transitions. In the doped systems, new transition pathways emerge involving the alkali metal $s$-orbitals (Na $3s$, K $4s$, and Rb $5s$). The progressive involvement of these larger, more diffuse orbitals in the Rb and K systems likely contributes to the more effective light absorption, underscoring the critical role of the dopant-induced local electronic states. This distortion-dominated enhancement of ultraviolet absorption is a critical factor for improving the material's sensitivity and overall efficiency in radiation dosimetry applications.

In summary, multi-scale simulations are employed to elucidate the microscopic mechanism by which alkali metal (Na, K, Rb) doping enhances the dosimetric properties of $LiMgPO_4$. We find that co-doping with all three alkali metals (Na, K, and Rb) expands the unit cell and narrows the electronic band gap from 5.47 eV in pristine $LiMgPO_4$ to 4.78 eV in the triply-doped system. Consequently, the absorption spectrum exhibits a significant redshift, and the effective band structure reveals new, dopant-induced localized states near the CBM. These computational trends, monotonically expanding unit cell and red-shifted absorption, mirror the experimental observations for Na-doped $LiMgPO_4$, confirming that the present mechanism-based picture captures the essential physics.

Analysis of the carrier dynamics reveals a crucial dichotomy. Doping suppresses electron mobility near the CBM at the Γ-point, while substantially enhancing hole mobility near the VBM. This behavior is governed by a synergy between the dopant-induced trap states and associated lattice distortions. Specifically, pronounced local lattice distortions surrounding the dopant sites promote electron self-trapping - the formation of small polarons - which effectively localizes the charge and stabilizes the stored energy. Conversely, the global strain across the crystal lattice facilitates hole delocalization and migration, thereby suppressing detrimental electron-hole recombination.

This work clarifies, at an atomic level, the key physical mechanism underpinning the enhanced performance of alkali-doped $LiMgPO_4$: a cooperative effect wherein localized distortions trap electrons for signal stability, while global distortions accelerate holes to prevent signal loss. Our findings address critical gaps in previous theoretical models, which have largely overlooked the roles of deep-level traps and lattice distortions. We thus establish a comprehensive microscopic framework for understanding and engineering the carrier dynamics in this class of radiation detection materials.

**COMPUTATIONAL METHODS**

First-principles calculations are performed using the Vienna *Ab initio* Simulation Package (VASP) [28,29] within the framework of spin-polarized density functional theory

(DFT). The projector-augmented wave (PAW) method is employed to describe the interaction between core and valence electrons.[30] Electron exchange and correlation effects were treated using the Perdew-Burke-Ernzerhof (PBE) functional within the generalized gradient approximation (GGA).[31] The PBE functional is chosen as it provides a reliable balance between computational efficiency and accuracy for predicting the geometric and electronic structures of oxide compounds such as LiMgPO$_4$, particularly for systems involving metal-oxygen bonds. We acknowledge that PBE is known to systematically underestimate the band gaps of wide-gap insulators. Our calculated pristine gap of 5.47 eV, for instance, is considerably lower than the experimental optical gap of 7.5 eV.[25] However, the primary conclusions of this study are not dependent on the absolute band gap value. Instead, they are derived from relative energy trends, structural distortions, defect formation energies, and non-adiabatic carrier dynamics. These properties, which rely on the accurate description of interatomic forces and the relative alignment of energy levels, are reliably captured at the PBE level. Therefore, PBE is considered an adequate and computationally feasible choice for elucidating the mechanistic trends at the core of our investigation.

A plane-wave cutoff energy of 500 eV is used for all calculations. The electronic self-consistency loop is converged to a tolerance of $10^{-6}$ eV. For structural relaxations, the geometry is optimized until the residual forces on all ions are less than 0.015 eV/Å. The Brillouin zone is sampled using a Γ-centered Monkhorst-Pack k-point mesh. To analyze the influence of defects on the electronic structure, the effective band structure (EBS) of each defect-containing supercell is calculated and unfolded onto the Brillouin zone of the primitive cell for direct comparison.[32,33] The EBS is calculated using a projection method, the methodology of which is detailed in Section S2 of the Supplemental Material. Optical properties are calculated within the independent particle approximation (IPA), based on the electronic structure obtained at the PBE level.[34] We note that the IPA neglects electron-hole interactions (i.e., excitonic effects), which can be significant in wide-gap ionic materials and typically lead to a redshift of the absorption onset. While this affects the absolute energy of the calculated absorption edge, this approach is sufficient for capturing the qualitative trends and relative shifts induced by doping, which are the focus of our study.

Furthermore, carrier dynamics are investigated using *ab initio* NAMD simulations as implemented in the Hefei-NAMD code.[35] This code interfaces with VASP and employs the fewest-switches surface hopping (FSSH) algorithm[36-38] to model nonadiabatic transitions. The dynamics are propagated by solving the time-dependent Kohn-Sham (TDKS) equations under the classical path approximation.[39-43] For each system (pristine, Na-, K-, and Rb-doped), the nonadiabatic dynamics are averaged over 100 independent trajectories to ensure statistical convergence. Each trajectory is initiated from a unique structural snapshot randomly sampled from the last 20 ps of the AIMD simulation at 300 K, ensuring a thorough sampling of the initial phase space.

## ACKNOWLEDGMENTS

This work is sponsored by the Natural Science Foundation of Xinjiang Uygur Autonomous Region (Grant No. 2023D01D03, 2022D01C689, and 2022D01C48), the

Xinjiang University Outstanding Graduate Student Innovation Project (Grant No. XJDX2025YJS040).

**SUPPORTING INFORMATION AVAILABLE**

Supplemental material including doping site selection and ab initio molecular dynamics (AIMD) simulations validating thermodynamic stability; defect formation energies under various chemical potentials; effective band structures and electron localization function (ELF) analyses elucidating carrier localization; additional non-adiabatic coupling data; and comprehensive optical absorption spectra for all doped and defective systems.

**AUTHOR DECLARATIONS**

The authors have no conflicts to disclose.

**DATE AVAILABILITY**

The data that support the findings of this study are available from the corresponding author upon reasonable request.

**FIGURES**:

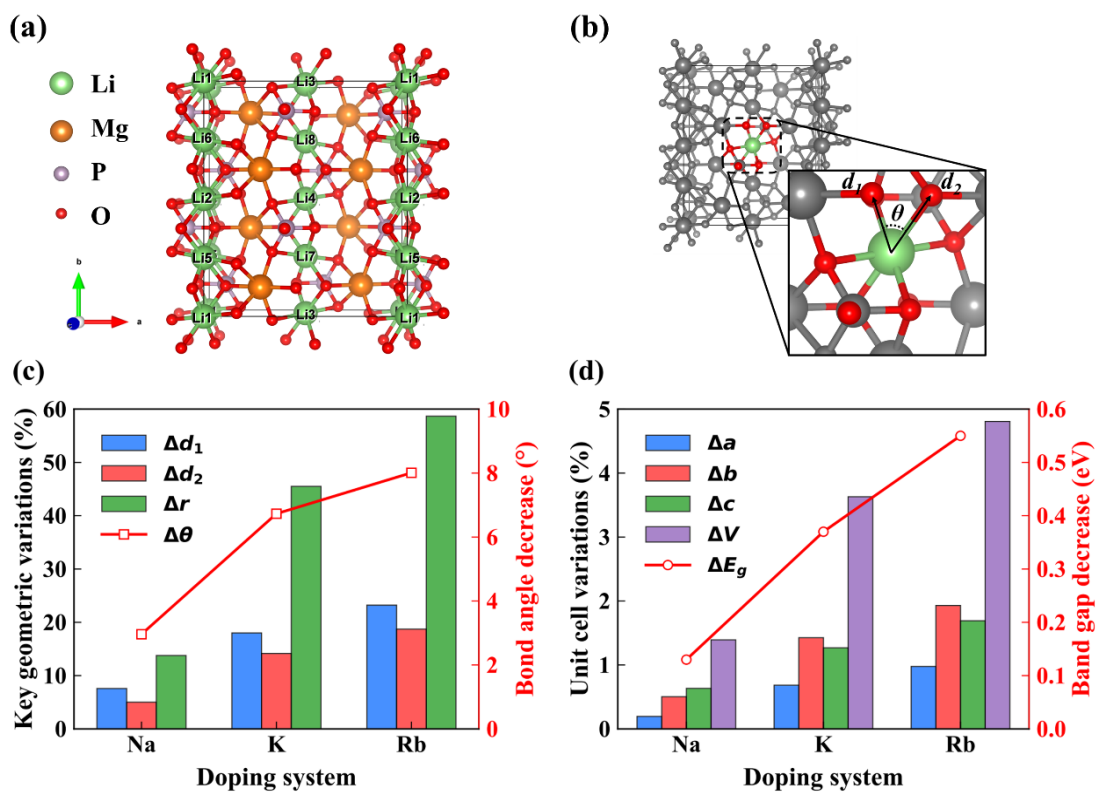

Figure 1. Structural and electronic consequences of alkali-metal doping in LiMgPO$_4$. (a) The pristine LiMgPO$_4$ computational supercell, with atoms identified by the legend. (b) Local atomic structure around a substitutional dopant at a Li site, defining the characteristic bond lengths ($d_1$, $d_2$) and bond angle ($\theta$). (c) Percentage changes in bond length ($\Delta d_1$, $\Delta d_2$) and ionic radius ($\Delta r$) are shown as bars (left axis), while the decrease in bond angle ($\Delta \theta$) is plotted as a line (right axis). (d) Percentage changes in lattice parameters ($\Delta a$, $\Delta b$, $\Delta c$) and volume ($\Delta V$) are shown as bars (left axis), while the corresponding decrease in band gap ($\Delta E_g$) is plotted as a line (right axis).

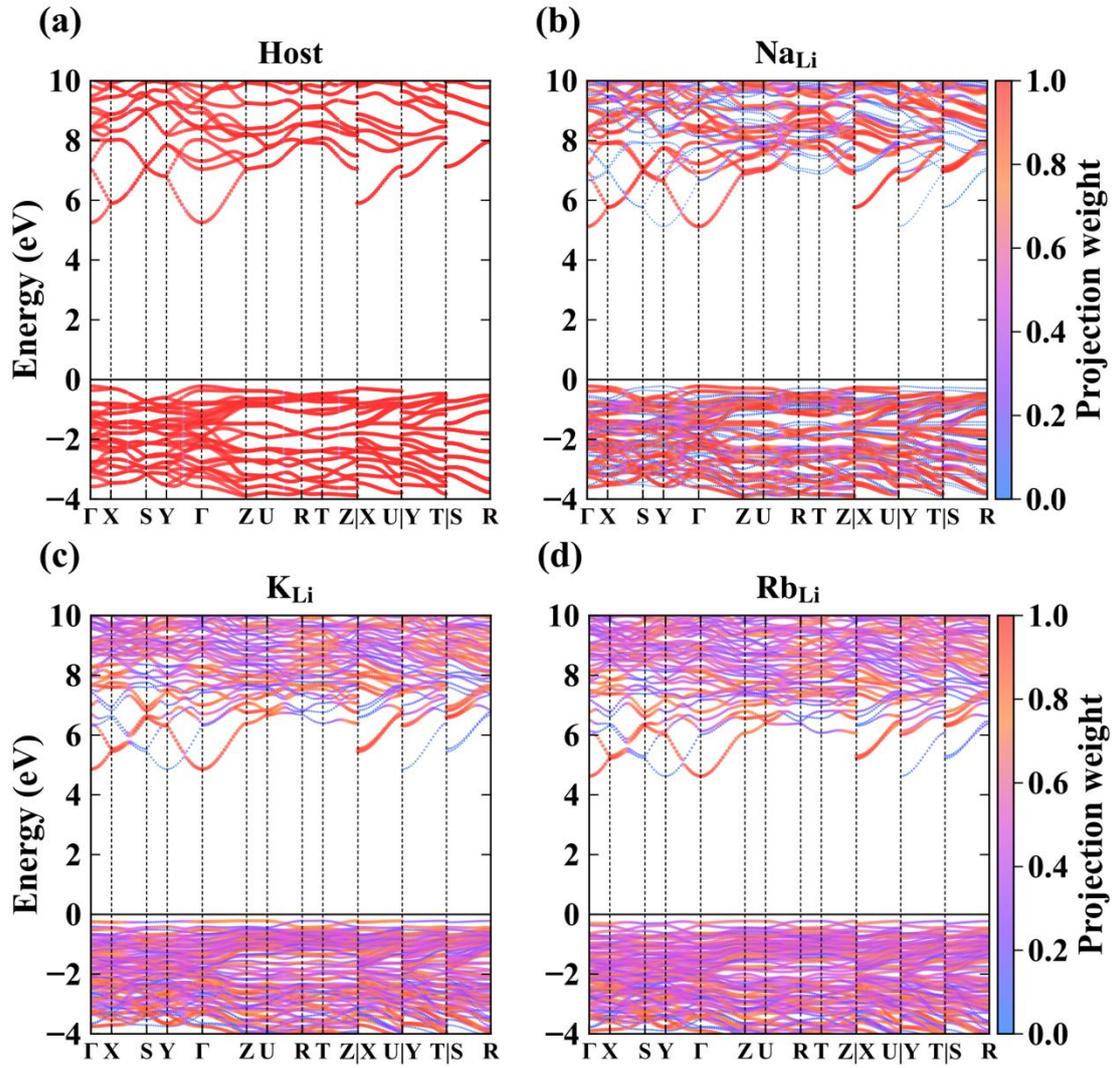

Figure 2. The effect of single-atom doping on the electronic band structure of LiMgPO$_4$, visualized using the effective band structure (EBS) method. The panels show the pristine host (a) in comparison to systems with Li sites substitutionally doped with (b) Na, (c) K, and (d) Rb. The valence band maximum (VBM) is set to 0 eV. The color mapping quantifies the projection weight (from 0.0 in blue to 1.0 in red), indicating the degree to which the electronic states retain the character of the pristine host's Bloch states.

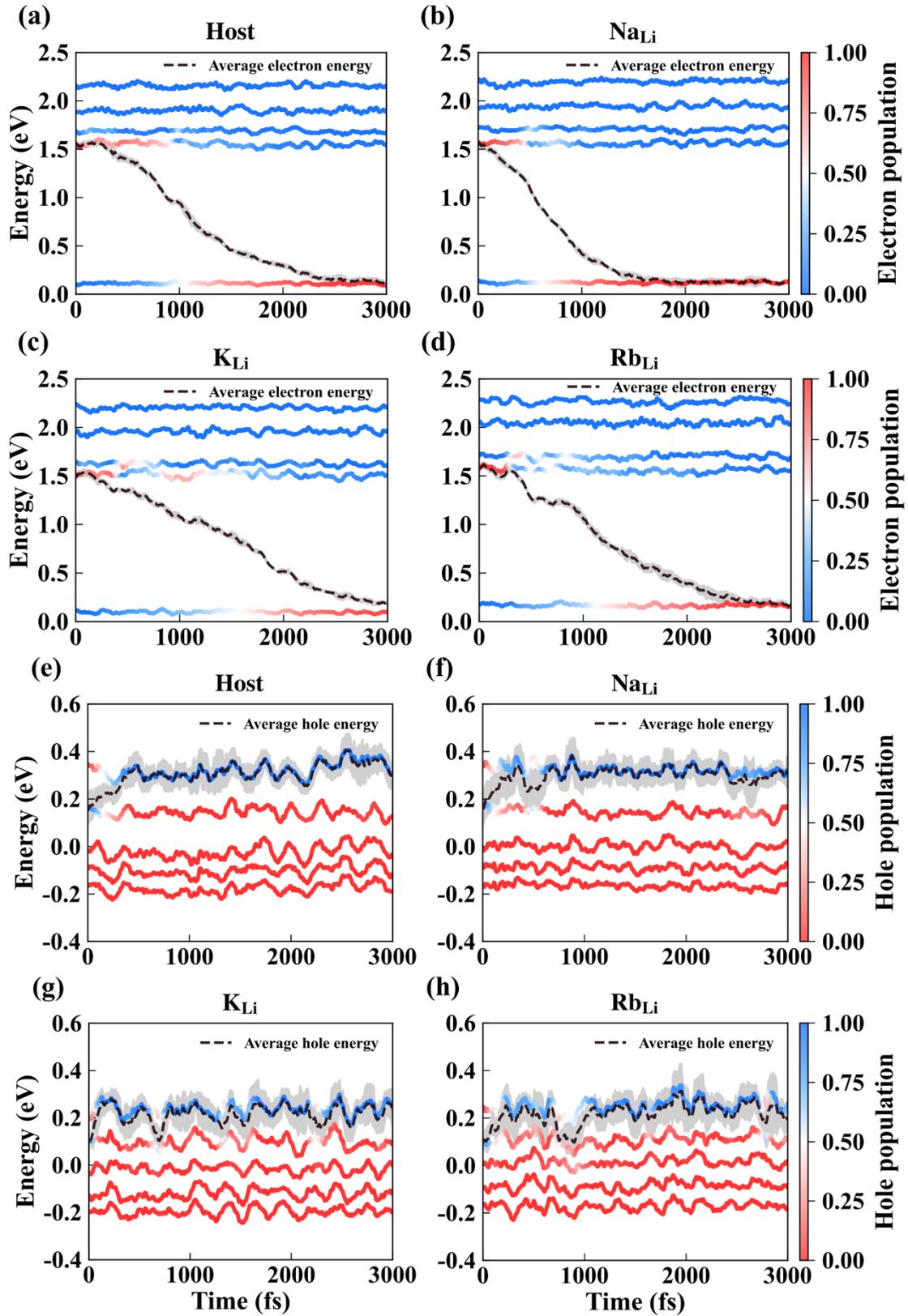

Figure 3. Charge carrier relaxation dynamics in pristine and doped LiMgPO$_4$ at 300 K over 3 ps. The panels (a)-(d) show the time-evolution of the ensemble-averaged electron energy distribution for (a) the undoped host, and systems doped with (b) Na, (c) K, and (d) Rb. The panels (e)-(h) display the corresponding hole energy dynamics for (e) the undoped host, and systems doped with (f) Na, (g) K, and (h) Rb. The color map illustrates the population density of electrons at each energy level. The black dashed line tracks the relaxation of the mean electron energy, averaged over 100 trajectories. The surrounding shaded region represents the standard error of the mean (SEM), indicating the statistical uncertainty of the relaxation pathway.

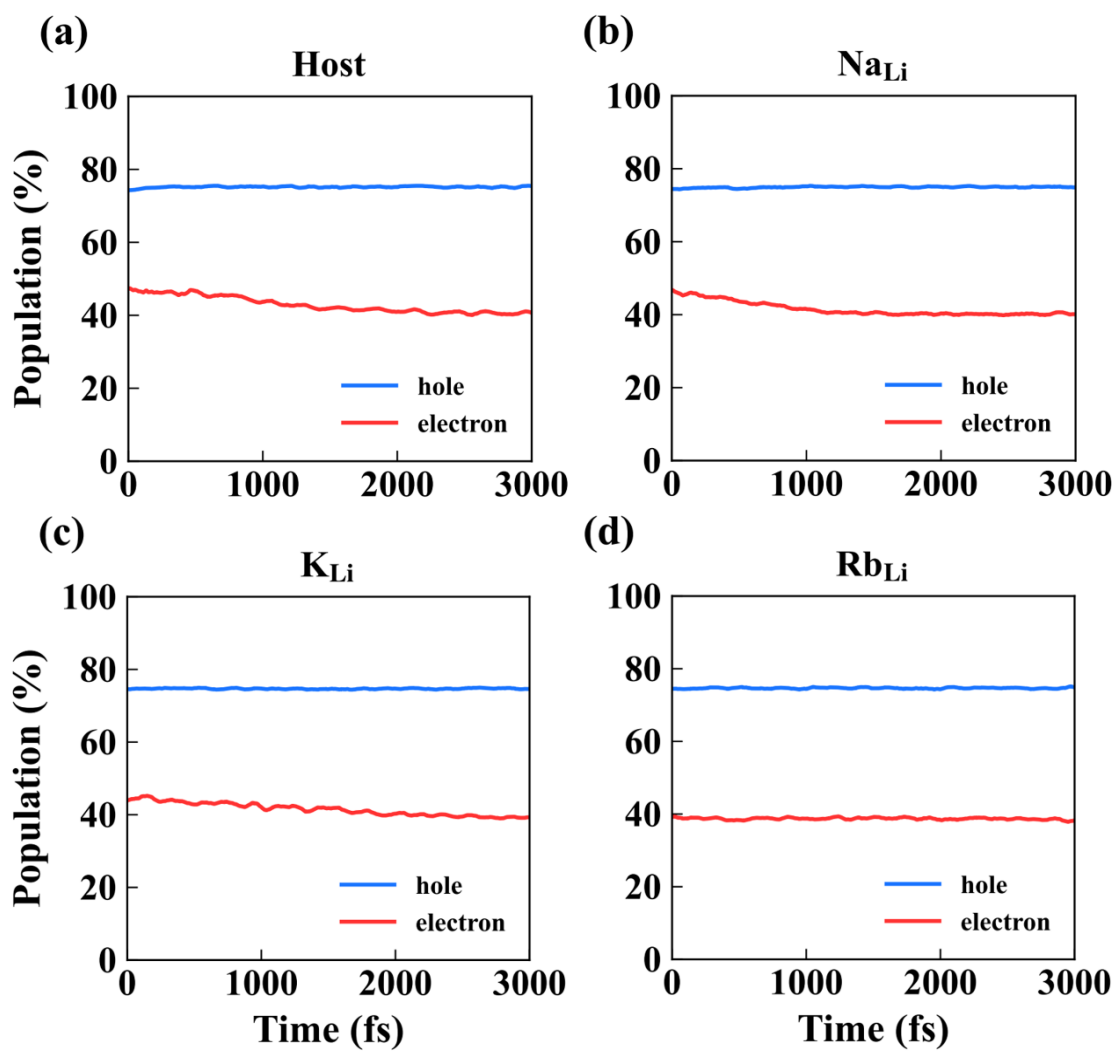

Figure 4. Quantifying charge carrier localization dynamics via non-adiabatic molecular dynamics. The figure shows the time evolution of the localization percentage for an excited electron (red) and hole (blue). The dynamics are compared for (a) the pristine host material and for hosts substitutionally doped at the Li site with (b) Na, (c) K, and (d) Rb. The stability of the localization is tracked over a 3 ps simulation time. The localization percentage is calculated by projecting the carrier wavefunction onto a dopant-centered defect fragment which consists of the substitutional dopant atom and its six nearest-neighbor oxygen atoms.

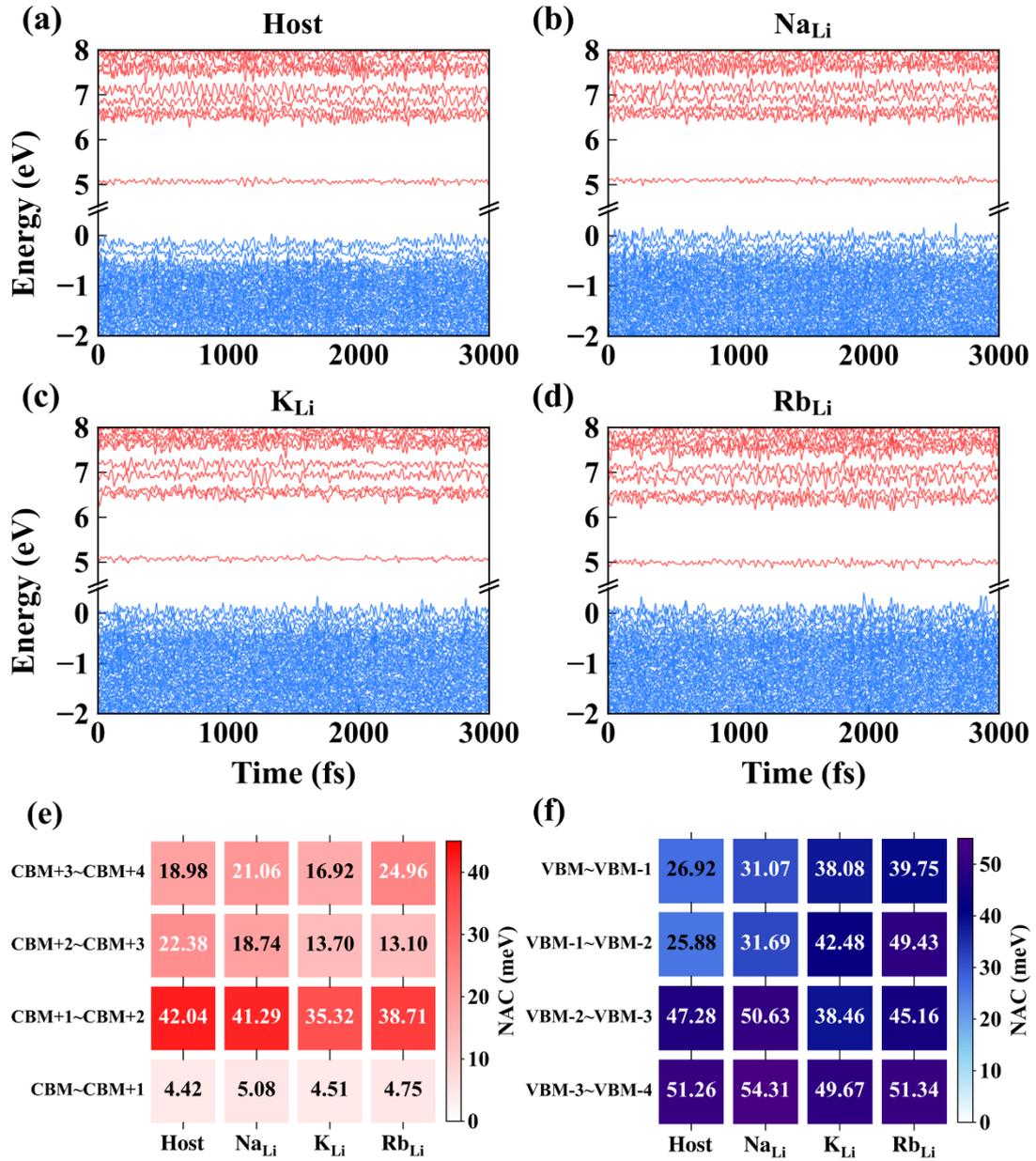

Figure 5. Evolution of Kohn-Sham (KS) orbital energies at 300 K over 3 ps and time-averaged root-mean-square non-adiabatic coupling (NAC) values. The panels depict (a) the pristine host and systems doped at the Li site with (b) Na, (c) K, and (d) Rb. All energies are aligned to the VBM of the initial pristine system. Conduction band (CB) and valence band (VB) states are shown in red and blue, respectively. Additionally, (e) shows the NAC values between the conduction band minimum (CBM) and CBM+4, and (f) shows the NAC values between the VBM and VBM-4. Values are given in meV for the pristine and doped systems.

**TABLES:**

Table 1. Structural and electronic parameters for pristine LiMgPO$_4$ and various defective configurations. Tabulated values include the absolute band gap ($E_g$) in eV, along with the percentage changes in supercell volume ($\Delta V/V_0$) and band gap ($\Delta E_g$). The defects considered are an intrinsic Li vacancy and extrinsic doping with single (Na, K, Rb) and multiple alkali metals. All relative changes are calculated with respect to the pristine system.

| System | $\Delta V/V_0$ (%) | $E_g$ (eV) | $\Delta E_g$ (eV) |
|---|---|---|---|
| LiMgPO$_4$ | 0.00 | 5.47 | 0.0 |
| **Li vacancy** | **+0.45** | **5.52** | **+0.9** |
| Na-doped | +1.39 | 5.34 | -2.4 |
| K-doped | +3.62 | 5.10 | -6.8 |
| Rb-doped | +4.80 | 4.92 | -10.1 |
| Na-K-doped | +4.90 | 5.05 | -7.7 |
| Na-Rb-doped | +6.08 | 4.90 | -10.4 |
| K-Rb-doped | +12.77 | 5.00 | -8.6 |
| **Na-K-Rb-doped** | **+14.91** | **4.78** | **-12.6** |

# Supplemental Material
# Distortion-Driven Carrier Decoupling in Doped LiMgPO$_4$

Zhihua Zheng[1], Xiaolong Yao[1,4,*], Cailian Yu[1], Menghao Gao[1], Fangping Ouyang[1,2,3,*], Shiwu Gao[4,*]

[1]School of Physical Science and Technology, Xinjiang Key Laboratory of Solid-State Physics and Devices, Xinjiang University, Urumqi 830017, China

[2]School of Physics, Institute of Quantum Physics, Hunan Key Laboratory for Super-Microstructure and Ultrafast Process, and Hunan Key Laboratory of Nanophotonics and Devices, Central South University, Changsha 410083, China

[3]State Key Laboratory of Powder Metallurgy, and Powder Metallurgy Research Institute, Central South University, Changsha 410083, People's Republic of China

[4]Beijing Computational Science Research Center, Beijing 100193, China

[*]Email: xlyao@xju.edu.cn, ouyangfp06@tsinghua.org.cn, swgao@csrc.ac.cn

## 1. Doping sites and AIMD simulations

We performed systematic modeling of multi-site doping configurations (double doping and triple doping, Figure S1a), and stability site selection confirmed Li2/Li4 (double doping) and Li3/Li4/Li5 (triple doping) as the optimal configurations (energy-minimized structures, Table S1). Subsequent studies focused on the most stable configurations in these systems.

We utilize AIMD simulations to validate the thermodynamic stability of the Na, K, and Rb co-doped LiMgPO$_4$. The simulations are conducted using the NVT ensemble, maintaining the constant number of particles, volume, and temperature. The $1 \times 2 \times 2$ supercell was simulated at 300 K with a time step of 3 fs, over the total simulation time of 15 ps. Figure S1b shows the energy evolution of the Na, K, and Rb co-doped LiMgPO$_4$ over 15 ps at 300 K. Figure S1b also displays the supercell configuration at the initial and final states. The total energy converges quickly and oscillates around the straight line for the remainder of the simulation, indicating that the crystal structure remains stable throughout, no dissociation and no vacancies occurred. This suggests excellent thermodynamic stability of the Na, K, and Rb co-doped LiMgPO$_4$ crystal.

## 2. Defect formation energies in different LiMgPO$_4$ systems

To understand the formation of variously doped LiMgPO$_4$ under different synthetic conditions, we have calculated and investigated the defect formation energies to suggest some guidance for experimental synthesis pathways. The defect formation energy can be defined as the function of the chemical potential, the number of host atoms, and the number of defect atoms, as shown in the following equation:[1]

$$\Delta H_f = H_{tot}^{def} - \sum_H n_H \mu_H - \sum_I n_I \mu_I, \qquad (1)$$

where $H_{tot}^{def}$ is the total energy of the supercell containing the defect, $n_H$ is the number of host elements, and $\mu_H$ is the chemical potential of the host element. $n_I$ is the number of dopant elements, and $\mu_I$ is the chemical potential of the dopant element.

The lower the defect formation energy, the easier it is to form the defect thermodynamically. Under equilibrium conditions between LiMgPO$_4$ and its constituent elements (Li, Mg, P, and O), the sum of the chemical potentials of these elements equals the chemical potential of bulk LiMgPO$_4$: $\mu_{Li} + \mu_{Mg} + \mu_P + 4\mu_O = \mu_{LiMgPO_4(bulk)}$. Therefore, the defect formation energy of our system is defined as:

$$\Delta H_f = E_{tot}^{def} - E_{tot} + n_{Li}\mu_{Li} - n_I\mu_I. \qquad (2)$$

The chemical potential of each element must not exceed its bulk value, so: $\mu_{Li} \leq \mu_{Li(bulk)}$, $\mu_{Mg} \leq \mu_{Mg(bulk)}$, $\mu_P \leq \mu_{P(bulk)}$, $\mu_O \leq \mu_{O(gas)}$. The formation energy of LiMgPO$_4$ in the thermodynamically stable state is given by: $\Delta = \mu_{LiMgPO_4(bulk)} - \mu_{Li(bulk)} - \mu_{Mg(bulk)} - \mu_{P(bulk)} - 4\mu_{O(gas)}$, where $\mu_{Li(bulk)}$, $\mu_{Mg(bulk)}$, and $\mu_{P(bulk)}$ are calculated from their respective bulk phases, and $\mu_{O(gas)}$ is calculated for the O$_2$ molecule in the 20 Å × 20 Å × 20 Å cubic lattice and averaged over each atom. The chemical potentials of the substituting atoms are also obtained from their respective bulk phases. We calculate the formation energy to be $\Delta = -20.21$ eV. Finally, the following relationship is derived:[2-4]

$$\mu_{Li(bulk)} + \Delta \leq \mu_{Li} \leq \mu_{Li(bulk)}. \tag{3}$$

Figure S2 shows the defect formation energy as a function of Li chemical potential for all defect systems. In the Li-rich case, the defect formation energies of all systems are greater than zero, indicating that defects are more difficult to form thermodynamically, which is expected given the difficulty in generating the necessary vacancies in a Li-rich environment. The defect formation energies for Li vacancies are consistent with those calculated by P. Modak *et al*..[5] Comparing K and Rb single doping with Na, the defect formation energy is higher under both Li-rich and Li-poor conditions, suggesting that defects involving K and Rb are more difficult to produce than those involving Na, likely due to the larger atomic radii of K and Rb. In the double-doped case, the defect formation energy increases with the atomic radii of the dopants under Li-rich conditions and is higher than that of Na defects. The defect formation energy in the triple-doped case is higher than in all single-doped cases. Interestingly, as conditions shift from Li-rich to Li-poor, the defect formation energies for double and triple doping exhibit the opposite trend, indicating that double and triple doping are more likely to form than single doping as the chemical potential of Li decreases. This suggests that better crystallization of double- and triple-doped LiMgPO$_4$ systems can be achieved by controlling the Li concentration during material fabrication.

## 3. Effective band structure analysis

To elucidate the impact of dopant-induced lattice distortions on carrier dynamics - a phenomenon intimately linked to the formation of polarons - we calculate the effective band structure (EBS) for the doped systems. The EBS method unfolds the electronic bands of the supercell, originally computed within its corresponding small Brillouin zone (BZ), back onto the larger BZ of the primitive unit cell.[6-8] This projection is essential for revealing how the electronic states of the pristine crystal are perturbed by the presence of the dopant.

The analysis is achieved by calculating the spectral weight, $P_{m,K}(k)$, which projects each supercell eigenstate $|\Psi_{m,K}\rangle$, (with band index $m$ and wavevector $K$) onto the complete set of Bloch states $|\psi_{m,k}\rangle$ (with band index $n$ and wavevector $k$) of the primitive cell. The wavevectors are related by the standard band-folding condition $K = k - G$, where $G$ is a reciprocal lattice vector of the primitive cell. The spectral weight is given by

$$P_{m,K}(k) = \sum_n |\langle |\psi_{m,k}||\Psi_{m,K}\rangle|^2. \tag{4}$$

Here, $P_{m,K}(k)$ quantifies the contribution of the primitive-cell Bloch characters at wavevector $k$ to the supercell eigenstate $m$. A high spectral weight (i.e., $P \approx 1$) at a specific $k$ signifies that the corresponding supercell state retains a well-defined crystal momentum, characteristic of a delocalized Bloch wave. Conversely, a smearing of the spectral weight across multiple $k$-points indicates strong state mixing and carrier localization, a hallmark of small polaron formation.

## 4. Electron localization function for single-doped systems

The electron localization function (ELF) illustrates the degree of electron localization at different locations within three-dimensional real space. Higher ELF values in the given region indicate that electrons are more confined to that region, while lower values suggest that electrons are more likely to diffuse outward. The ELF is defined as:[9]

$$ELF = (1 + \chi_\sigma^2)^{-1}, \tag{5}$$

where $\chi_\sigma = D_\sigma/D_\sigma^0$. Here, $D_\sigma^0$ represents the localized description of the ideal uniform electron gas at the spin density equal to the local value of $\rho_\sigma(r)$, and $\chi_\sigma$ is the dimensionless localization index. $D_\sigma^0 = \left(\frac{3}{5}\right)(6\pi^2)^{2/3}\rho_\sigma^{5/3}$. Therefore, the ELF is bounded within the range:

$$0 \leq ELF \leq 1. \tag{6}$$

Figure S5a presents the schematic diagram of the ELF cuts, shaded in blue, where substitution atoms and the four nearest-neighbor O atoms were chosen to study their electronic distributions after bonding. Figures S5b-e display two-dimensional ELF plots for pristine LiMgPO$_4$ and Na-, K-, and Rb-doped LiMgPO$_4$, respectively. The investigation reveals that the electrons of the Li atom are highly localized in the nucleus, with localization approaching 1. In contrast, the electronic localization of Na atoms is extremely low, while the localization of K and Rb atoms tends to be closer to that of O atoms. Interestingly, the change in electronic localization of the dopant elements does not show a linear relationship with the atomic number progression within the same group. As the electronic localization of the dopant elements decreases, the localization of the O atoms also appears to decrease to some extent. It is noteworthy that O atoms exhibit the more pronounced lone-pair electron feature and combined with the DOS showing contributions near the top of the valence band originating from O, it is inferred that lone-pair electrons are one of the factors enhancing the material's optical properties. The decrease in the electron localization of O atoms due to the lower localization of the doped atoms compared to Li leads to the production of stronger lone-pair electrons in the O atoms, which results in improved light absorption performance. Additionally, there is no hybridization between the dopant and O atoms, and the electrons do not contribute to bonding.

## 5. Optical absorption properties of pristine and doped LiMgPO$_4$

We calculate the frequency-dependent dielectric function using linear response theory to derive the UV absorption spectrum of the LiMgPO$_4$ material. The complex dielectric function, $\varepsilon(\omega)$, is defined as $\varepsilon(\omega) = \varepsilon_1(\omega) + i\varepsilon_2(\omega)$, where $\varepsilon_1(\omega)$ and $\varepsilon_2(\omega)$ are the real and imaginary parts of the dielectric function, respectively, and $\omega$ is the photon frequency. The imaginary part is obtained from optical transitions from the valence to conduction states, while the real part is derived using the Kramers-Kronig relation. The relationship between the absorption coefficient and the real and imaginary parts of the dielectric function is given by:[8]

$$\alpha(\omega) = \frac{\sqrt{2}\omega}{c}\left(\sqrt{\varepsilon_1^1 + \varepsilon_2^2} - \varepsilon_1\right)^2 \quad (7)$$

Figure S8 show the absorption spectra of the single-doped, double-doped and triple-doped systems, respectively. In Figure S8b, all three double-doping configurations enhance the absorption, outperforming pristine LiMgPO$_4$. Notably, the band gaps for the Na, Rb, and K, Rb co-doped systems are 4.90 eV and 5.00 eV, respectively, with both showing similar levels of performance enhancement. There is no observed correlation between the number of atoms and the degree of improvement. Based on the effective band structure, we infer that the newly introduced energy levels near the bottom of the conduction band compensate for the bandgap reduction, leading to similar performance in these systems. In Figure S8c, we observe that compared to pristine LiMgPO$_4$, the presence of the Li vacancy leads to the disappearance of several characteristic peaks between 160 nm and 190 nm, consistent with the bandgap increase. And the absorption spectrum for the Na, K, and Rb co-doped system, where a significant enhancement is observed in the UV region. The DOS evaluation confirms that orbitals provided by the dopant elements facilitate easier transitions, contributing to the observed performance boost. Overall, the absorption coefficients of all systems, except for the Li vacancy system, shift toward the near-UV region, resulting in improved optical performance across the board.

**FIGURES:**

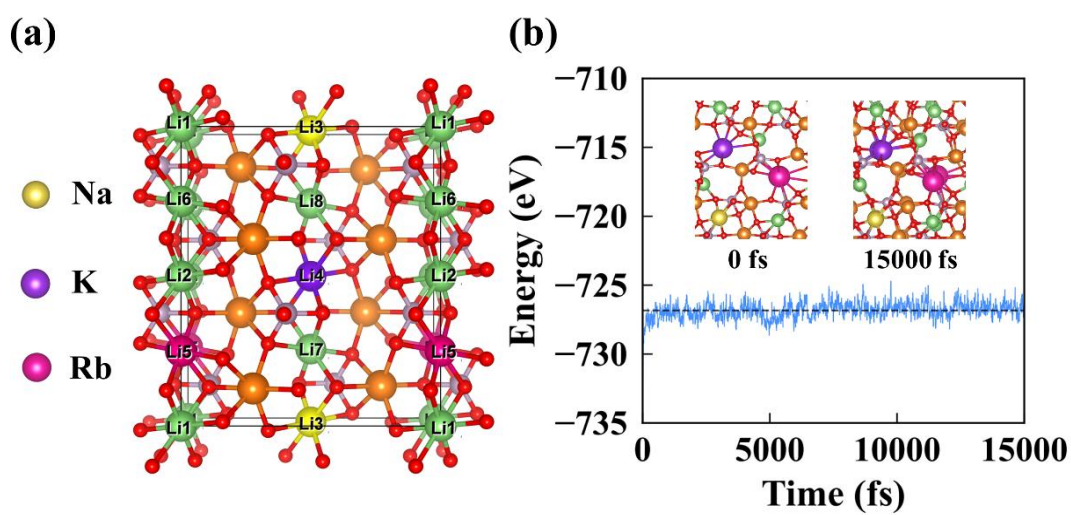

Figure S1. Stability of the Na-K-Rb co-doped LiMgPO$_4$ structure at 300 K. (a) The computational supercell used for the simulation. (b) Energy evolution during a 15,000 fs AIMD simulation. The minimal fluctuation in total energy, combined with the structural integrity maintained between the initial (0 fs, inset left) and final (15,000 fs, inset right) configurations, demonstrates the system's thermal stability.

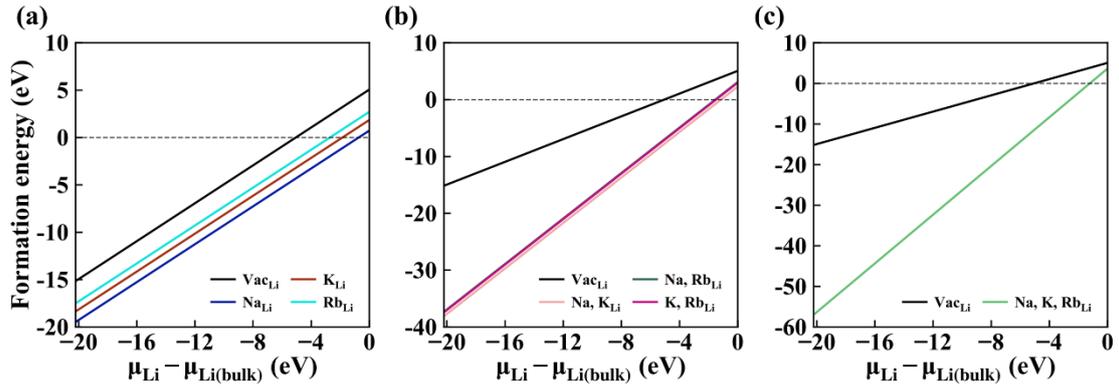

Figure S2. Defect formation energy diagrams comparing an intrinsic Li vacancy (Vac$_{Li}$, black line) with various extrinsic dopant configurations at the Li site. The plots are shown as a function of the Li chemical potential for (a) single-doping, (b) double-doping, and (c) triple-doping scenarios. Across all thermodynamically relevant conditions, the substitutional dopants exhibit significantly lower formation energies than the Li vacancy, demonstrating their energetic preference.

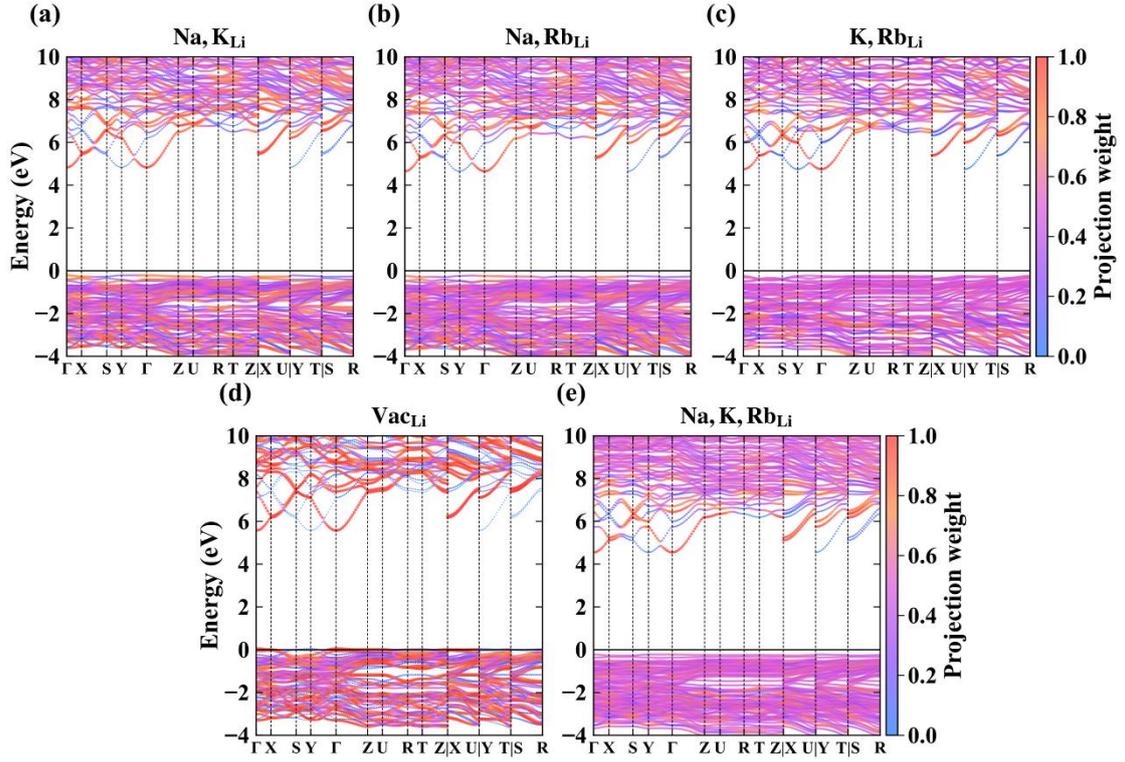

Figure S3. Effective band structures for various co-doped and defect systems. The panels correspond to substitutions at the Li site: (a) Na, $K_{Li}$, (b) Na, $Rb_{Li}$, (c) K, $Rb_{Li}$, (d) $Vac_{Li}$, and (e) Na, K, $Rb_{Li}$. In all plots, the energy scale is referenced to the Valence Band Maximum (VBM) at 0 eV. The colors in the effective band structure reflect the projection weights of the original single-cell Bloch states at the corresponding k points: red indicates high weight (dominant), and blue indicates low weight (small contribution).

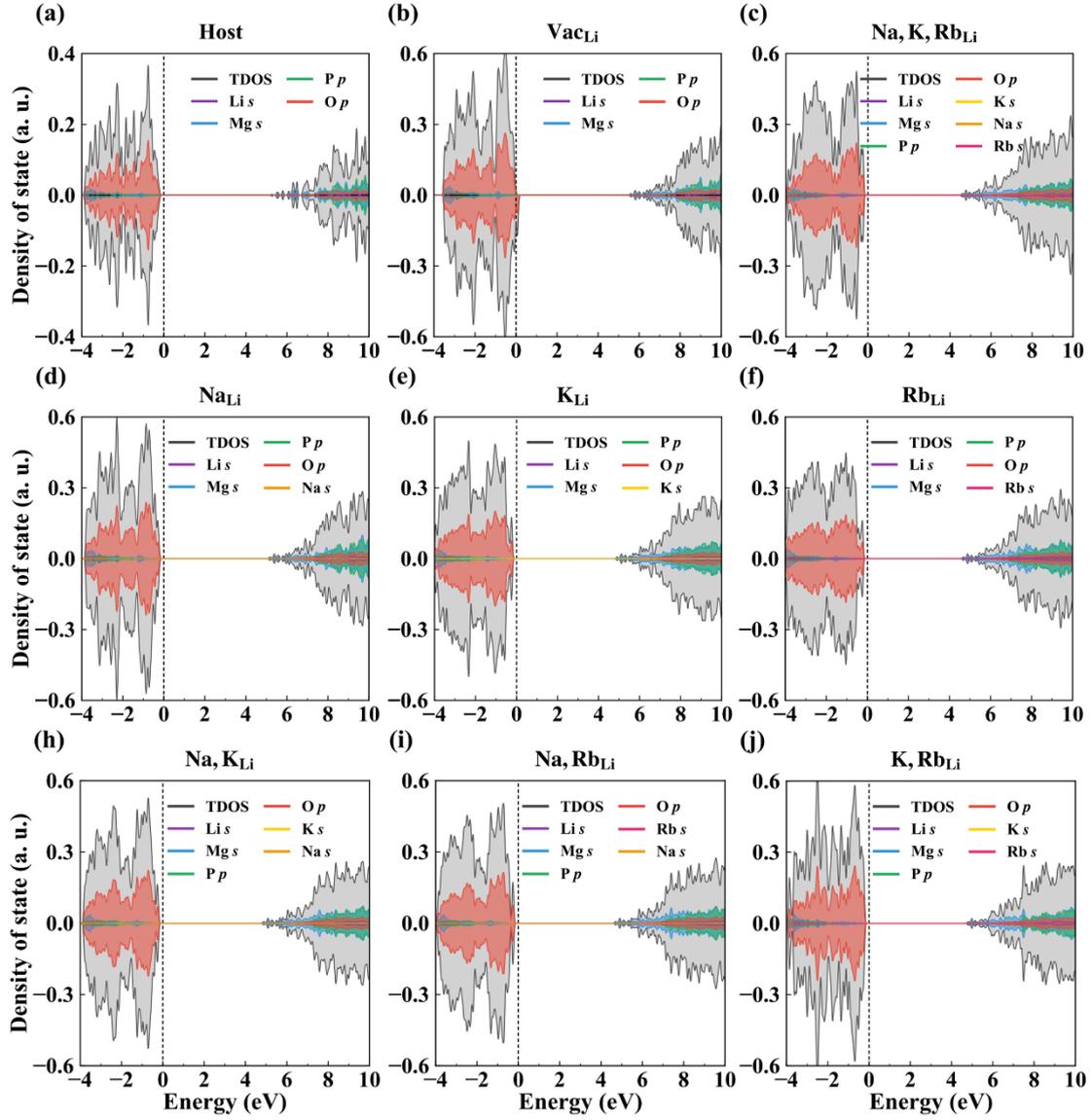

Figure S4. Total and orbital-projected density of states (DOS) for the host material and various Li-site defect configurations. Panels show: (a) the pristine host, (b) $Vac_{Li}$, (c) the triply-doped Na, K, $Rb_{Li}$ system, (d-f) singly-doped systems ($Na_{Li}$, $K_{Li}$, $Rb_{Li}$), and (g-i) co-doped systems (Na, $K_{Li}$; Na, $Rb_{Li}$; K, $Rb_{Li}$). The total DOS (TDOS) is shown as a grey shaded area, with colored lines representing the projected DOS from constituent atomic orbitals. The energy scale is referenced to the Fermi level (0 eV).

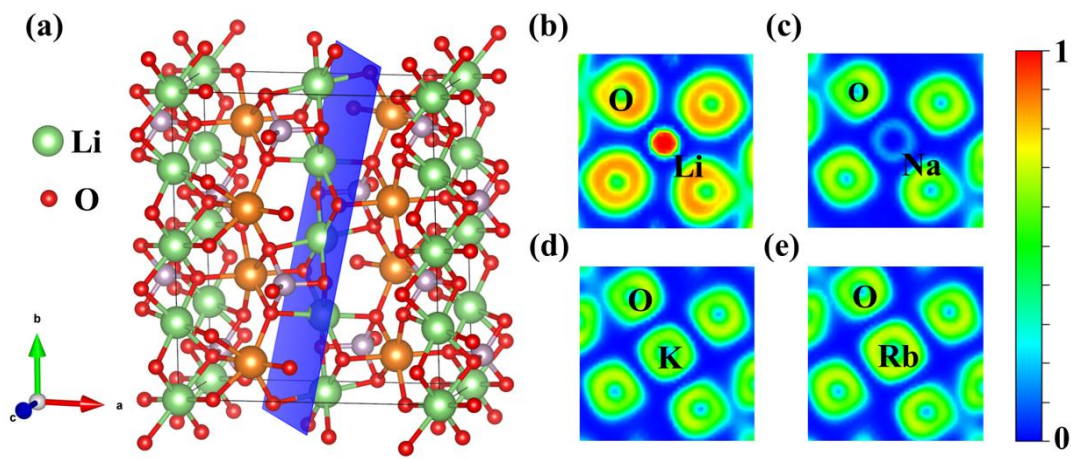

Figure S5. Visualization of chemical bonding via the Electron Localization Function (ELF). (a) The host material's crystal structure (Li: green, O: red), with a representative slicing plane shown in blue. Corresponding 2D ELF maps for the (b) pristine system and systems substitutionally doped with (c) Na, (d) K, and (e) at the Li site. The color scale quantifies the localization: red (ELF = 1) indicates high localization, such as in covalent bonds or lone pairs, while blue (ELF = 0) indicates delocalized, metallic-like regions.

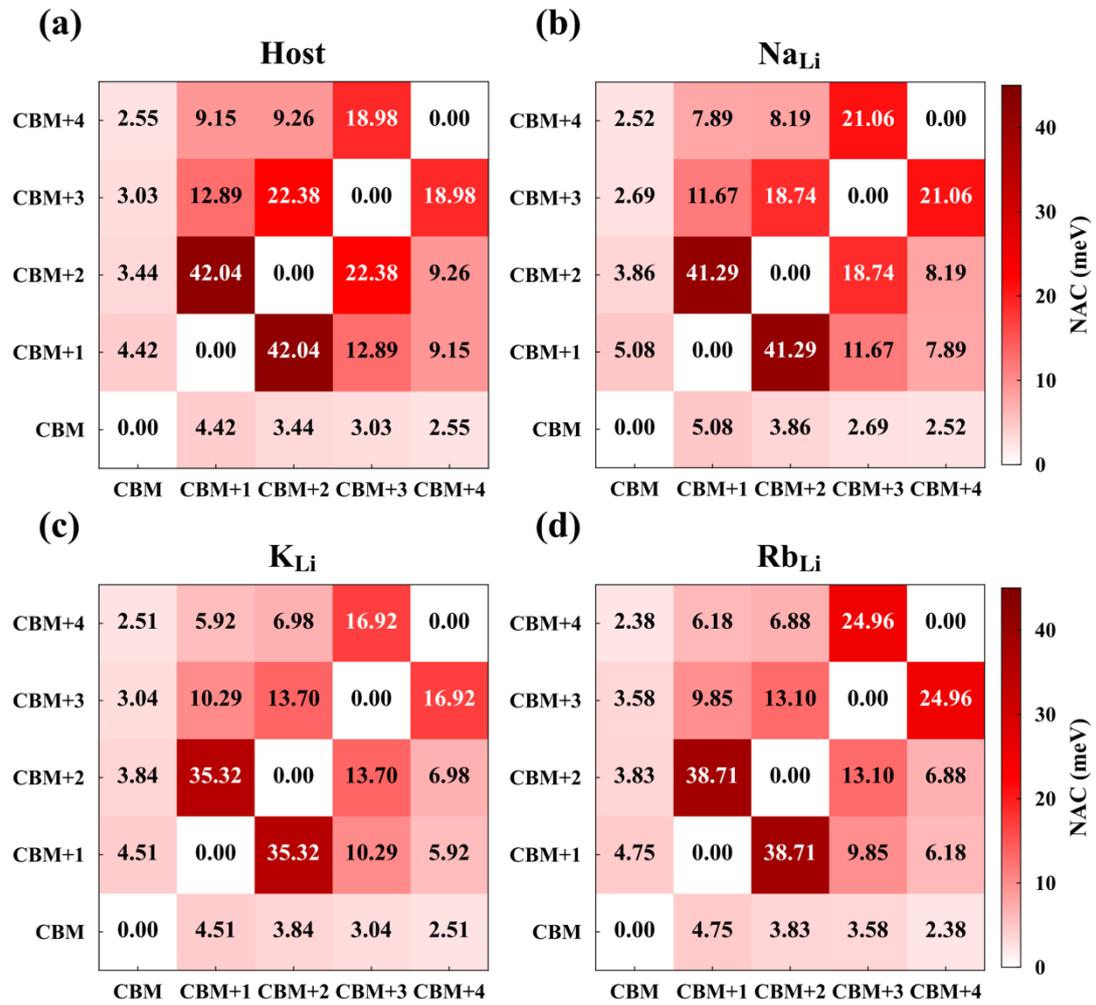

Figure S6. Time-averaged root-mean-square non-adiabatic coupling (NAC) values. The figure shows the (a) pristine and systems doped with (b) Na, (c) K, and (d) Rb at the Li site. The NAC calculation range is from the conduction band minimum (CBM) to CMB+4. Values are given in meV for the pristine and doped systems.

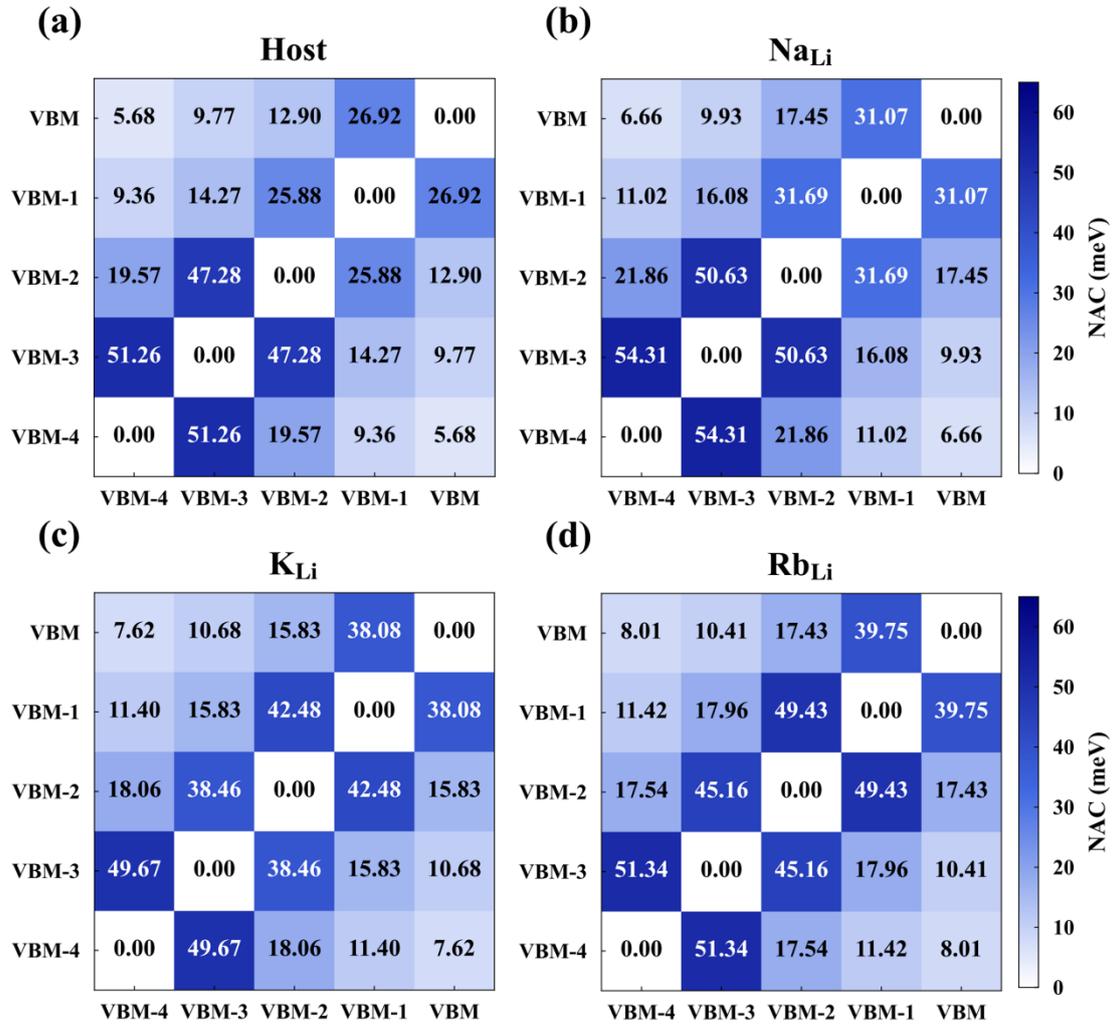

Figure S7. Time-averaged root-mean-square non-adiabatic coupling (NAC) values. The figure shows the (a) pristine and systems doped with (b) Na, (c) K, and (d) Rb at the Li site. The NAC calculation range is from the valence band maximum (VBM) to VBM-4. Values are given in meV for the pristine and doped systems.

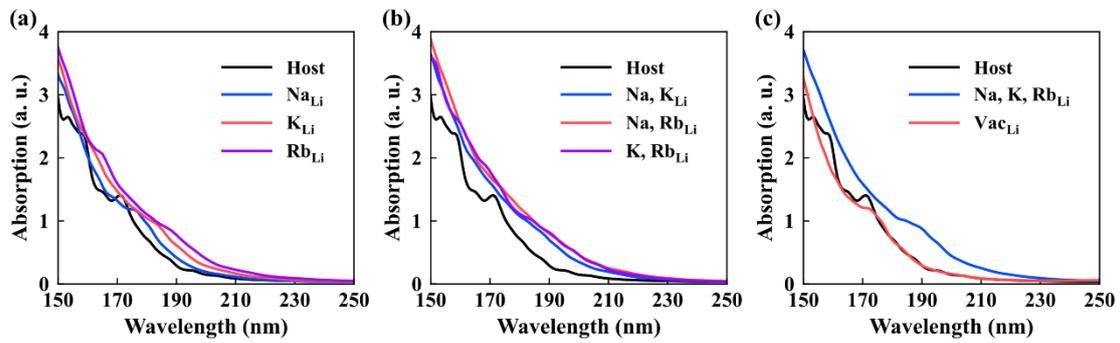

Figure S8. The effect of substitutional doping and vacancies on the optical absorption spectrum. The absorption of the pristine host material (black curve) is compared against: (a) single doping at the Li-site with Na, K, and Rb; (b) co-doping at the Li-site with Na/K, Na/Rb, and K/Rb pairs; and (c) tri-doping (Na, K, $Rb_{Li}$) and $Vac_{Li}$. A notable redshift in the absorption edge is observed for all modified systems, indicating a reduction in the optical band gap.

**TABLES:**

Table S1. Site preference analysis and total energies (eV) for various doped configurations. To determine the most stable substitution sites, the total energy of the optimized configurations was calculated for two identical Na, K, or Rb dopant atoms at non-equivalent double Li sites (double doping) and for Na, K, and Rb dopant atoms at non-equivalent triple Li sites (triple doping). For each series of calculations involving each dopant element (Na, K, Rb) at each doping level (double doping, triple doping), the underlined energy corresponds to the ground state (i.e., the most stable) configuration.

| Site | Li1-Li4(Na) | <u>Li2-Li4(Na)</u> | Li3-Li4(Na) | Li4-Li5(Na) | Li4-Li7(Na) |
|---|---|---|---|---|---|
| Energy (eV) | -369.97 | <u>-370.08</u> | -370.01 | -370.01 | -369.99 |
| Site | Li1-Li4(K) | <u>Li2-Li4(K)</u> | Li3-Li4(K) | Li4-Li5(K) | Li4-Li7(K) |
| Energy (eV) | -367.30 | <u>-367.86</u> | -367.39 | -367.46 | -367.40 |
| Site | Li1-Li4(Rb) | <u>Li2-Li4(Rb)</u> | Li3-Li4(Rb) | Li4-Li5(Rb) | Li4-Li7(Rb) |
| Energy (eV) | -365.59 | <u>-367.51</u> | -365.68 | -367.24 | -365.65 |
| Site | Li1-Li2-Li4 | Li3-Li4-Li7 | <u>Li3-Li4-Li5</u> | Li1-Li4-Li5 | Li2-Li4-Li5 |
| Energy (eV) | -366.45 | -365.49 | <u>-366.51</u> | -366.39 | -366.23 |

Table S2. Summary of calculated structural parameters and electronic properties for all simulated LiMgPO$_4$ systems. The table lists the optimized lattice constants ($a$, $b$, $c$ in Å), cell volume (in Å$^3$), and band gap ($E_g$ in eV) for the pristine host, a Li-vacancy defect, and various single, co-, and tri-doped configurations where dopants substitute for Li.

| System | $a$, $b$, $c$ (Å) | Volume (Å$^3$) | $E_g$ (eV) |
|---|---|---|---|
| LiMgPO$_4$ | 10.24, 11.92, 4.73 | 576.82 | 5.47 |
| LiMgPO$_4$: Li$_{vac}$ | 10.24, 11.90, 4.76 | 579.44 | 5.52 |
| LiMgPO$_4$: Na | 10.26, 11.98, 4.76 | 584.85 | 5.34 |
| LiMgPO$_4$: K | 10.31, 12.09, 4.79 | 597.76 | 5.10 |
| LiMgPO$_4$: Rb | 10.34, 12.15, 4.81 | 604.55 | 4.92 |
| LiMgPO$_4$: Na, K | 10.31, 12.17, 4.82 | 605.19 | 5.05 |
| LiMgPO$_4$: Na, Rb | 10.33, 12.24, 4.84 | 611.89 | 4.90 |
| LiMgPO$_4$: K, Rb | 10.17, 13.35, 4.79 | 650.50 | 5.00 |
| LiMgPO$_4$: Na, K, Rb | 10.22, 13.23, 4.90 | 662.91 | 4.78 |

Table S3. Local structural distortion around the Li-site induced by single-atom substitutional doping. The table details the change in key geometric parameters: the bond lengths between the substitutional atom (Li, Na, K, or Rb) and its nearest-neighbor oxygens ($d_1$, $d_2$), the associated O-dopant-O bond angle ($\theta$), and the corresponding atomic radius ($r$) of the element, included for reference.

| system | $LiMgPO_4$ | $LiMgPO_4$: Na | $LiMgPO_4$: K | $LiMgPO_4$: Rb |
|---|---|---|---|---|
| $d_1$, $d_2$ (Å) | 2.11, 2.19 | 2.27, 2.30 | 2.49, 2.50 | 2.60, 2.60 |
| $\theta$ (°) | 90.12 | 87.16 | 83.39 | 82.11 |
| $r$ (pm) | 167 | 190 | 243 | 265 |